\documentclass[acmsmall,authorversion]{acmart}
\usepackage{array}
\AtBeginDocument{%
  \providecommand\BibTeX{{%
    \normalfont B\kern-0.5em{\scshape i\kern-0.25em b}\kern-0.8em\TeX}}}

\setcopyright{acmcopyright}
\acmJournal{PACMCGIT}
\acmYear{2021} 
\acmVolume{4} 
\acmNumber{1} 
\acmArticle{} 
\acmMonth{5} 
\acmPrice{15.00}
\acmDOI{10.1145/3451255}

\begin{document}

\title{Virtual Reality Sickness Mitigation Methods: A Comparative Study in a Racing Game}


\author{Rongkai Shi}
\affiliation{%
  \institution{Xi'an Jiaotong-Liverpool University}
  \city{Suzhou}
  \state{Jiangsu Province}
  \country{China}}
\email{rongkai.shi19@student.xjtlu.edu.cn}

\author{Hai-Ning Liang}
\authornote{Corresponding author who can be contacted via haining.liang@xjtlu.edu.cn}
\affiliation{%
  \institution{Xi'an Jiaotong-Liverpool University}
  \city{Suzhou}
  \state{Jiangsu Province}
  \country{China}}
\email{haining.liang@xjtlu.edu.cn}

\author{Yu Wu}
\affiliation{%
  \institution{Xi'an Jiaotong-Liverpool University}
  \city{Suzhou}
  \state{Jiangsu Province}
  \country{China}}
\email{yu.wu18@student.xjtlu.edu.cn}

\author{Difeng Yu}
\affiliation{%
  \institution{University of Melbourne}
  \city{Melbourne}
  \state{Victoria}
  \country{Australia}}
\email{difengy@student.unimelb.edu.au}

\author{Wenge Xu}
\affiliation{%
  \institution{Xi'an Jiaotong-Liverpool University}
  \city{Suzhou}
  \state{Jiangsu Province}
  \country{China}}
\email{wenge.xu@xjtlu.edu.cn}


\begin{abstract}
Using virtual reality (VR) head-mounted displays (HMDs) can induce VR sickness. VR sickness can cause strong discomfort, decrease users’ presence and enjoyment, especially in games, shorten the duration of the VR experience, and can even pose health risks. Previous research has explored different VR sickness mitigation methods by adding visual effects or elements. Field of View (FOV) reduction, Depth of Field (DOF) blurring, and adding a rest frame into the virtual environment are examples of such methods. Although useful in some cases, they might result in information loss. This research is the first to compare VR sickness, presence, workload to complete a search task, and information loss of these three VR sickness mitigation methods in a racing game with two levels of control. To do this, we conducted a mixed factorial user study (N = 32) with degree of control as the between-subjects factor and the VR sickness mitigation techniques as the within-subjects factor. Participants were required to find targets with three difficulty levels while steering or not steering a car in a virtual environment. Our results show that there are no significant differences in VR sickness, presence and workload among these techniques under two levels of control in our VR racing game. We also found that changing FOV dynamically or using DOF blur effects would result in information loss while adding a target reticule as a rest frame would not.
\end{abstract}



\maketitle

\section{Introduction}
Virtual Reality (VR) Head-Mounted Displays (HMDs) provide immersive virtual environments (VEs) that have been used in many areas, particularly games. However, VR sickness, also known as cybersickness or virtual reality induced motion sickness (VRIMS), may occur with severe symptoms during or after users’ VR experience \cite{Re1,Re2}. Its symptoms include eye strain, headache, pallor, sweating, dryness of mouth, fullness of stomach, disorientation, vertigo, ataxia, nausea, and vomiting \cite{Re3}. These symptoms are similar as those from simulator and motion sickness but have different physiological causes \cite{Re3,Re4,Re5}. Motion sickness occurs with the disorder of the senses during travelling \cite{Re6} while simulator sickness is induced by illusionary visual stimulation in virtual simulators \cite{Re4,Re7,Re8}. The causes of VR sickness are still not completely understood, and common explanations include the poison theory, postural instability theory and sensory conflict theory \cite{Re9}. VR sickness appears quite common for people and the incidence is high. Sharples et al. \cite{Re1} found that 80\% of their over 200 participants experienced negative symptoms during VR exposure. In addition to health concerns, it may also result in shortened duration of the VR experience, decreased presence, enjoyment, or usefulness of the VE, and poor user performance in various environments especially in games \cite{Re10} which often requires movement and navigation activities. It is a main hindrance for the wider adoption of VR HMDs and greater production of VR content \cite{Re2}.

The nature of the tasks and how VEs are designed are two factors that influence VR sickness \cite{Re4,Re7,Re8,Re9,ReDN}. The interactivity between users and VEs is a determinant of the level of sickness \cite{Re11}. However, the effects of the degree of control in VEs appear to be largely unexplored, particularly using experiments. On the other hand, prior studies have explored and applied different VR sickness mitigation methods by adding visual effects or elements to VEs to help alleviate the effects of VR sickness. Manipulating the Field of View (FOV) is one well-known method (e.g., \cite{Re11,Re12,Re13,Re14,Re15,Re16,Re17}). FOV refers to the horizontal and vertical angular dimensions of the display \cite{Re7} and reducing it can increase viewing comfort. Another technique is based on changing the Depth of Field (DOF) blur \cite{Re18,Re19,Re20} to make some objects sharper (or blurrier) based on the distance between them and the users to help them focus on certain areas. Blurring can alleviate the accommodation-convergence visual conflict between objects of different distances to decrease visual discomfort. Additionally, some studies have found that providing a rest frame, a fixed, constant visual element in the display (such as a virtual nose \cite{Re17,Re21,Re22,Re23} or a target reticule \cite{Re24} in VEs), can help minimize VR sickness. A rest frame allows users to focus their gaze on it to reduce the incidence of visual information conflicts while moving virtually but not physically \cite{Re25}. Although these techniques are useful for mitigating VR sickness, they may result in information loss. Reducing the FOV decreases users' sense of presence in VEs \cite{Re11}, because the method blacks out certain parts of the user’s view. Likewise, using blur effects leaves only a portion of the view clearly visible. Similarly, providing a rest frame also block an area of users’ view. Despite their inherent weakness, there has been limited research that has looked into the information loss when using these techniques, especially relative to each other. 

In this paper, we evaluate in a systematic manner the relative effects among techniques for mitigating VR sickness, including adding a FOV restrictor, applying DOF blur effect, and attaching a target reticule as a fixed rest frame. We compared these methods with a baseline condition in terms of their effects on VR sickness, presence, workload to complete a search task, and information loss in a racing game with two degrees of control. 

\section{Related Work}
Several visual methods have been utilized to help reduce VR sickness. We focused on three techniques that have been widely discussed and applied: field of view reduction, depth of field blurring, and fixed rest frames. Other methods, such as adding visual path \cite{Re26}, applying dot effect \cite{Re27}, or changing the textures of VEs \cite{Re17}, blurring with the aid of deep learning \cite{Re28}, have not been examined thoroughly or faced difficulties in wide adoption. Table~\ref{tab:literature} provides a brief summary of the methods that are compared in our study. 

\begin{table*}
  \caption{Summary of the three visual reduction techniques for mitigating VR sickness that are compared in our study.}
  \label{tab:literature}
    \begin{tabular}{p{0.09\linewidth}p{0.15\linewidth}p{0.20\linewidth}p{0.26\linewidth}p{0.15\linewidth}l}
    \toprule
    Method&Description&Advantage&Disadvantage&Example of Usage Scenario\\
    \midrule
    FOV \newline Reduction&Reducing users' FOV&Lower incidence of sensory conflict&Blacks out part of the screen which may result in information loss&Spatial navigation \cite{Re13}\\
    \hline
    DOF Blur&Blurring of objects outside a pre-defined distance based on visual depth&Lower accommodation-convergence conflict&Blurred areas could affect performance and would not reduce the discomfort if the blur level does not match the depth level as in the physical reality&First-Person Shooting games \cite{Re20}\\
    \hline
    Rest Frame&Providing a fixed reference point of view in the VE&Helps focus users' eye gaze and reduce sensory mismatch&A rest frame always present in the view and can block other objects&Maze type of games \cite{Re34}\\
    \bottomrule
  \end{tabular}
\end{table*}

Prior studies have shown positive effects of restricting the Field of View (FOV) on lowering VR sickness \cite{Re11,Re12,Re13,Re14,Re17,Re22}. The sense of self-motion can be decreased by restricting the display’s FOV, which according to the sensory conflict theory can help avoid conflicts of senses, thereby helping to minimize sickness \cite{Re15}. Though restricting the FOV can reduce the level of sickness, users’ spatial degree of freedoms in VE is also decreased because of the increased non-visible part. This can lower the feeling of presence, enjoyment, and the performance in some tasks \cite{Re11,Re15,Re29}. To balance the tradeoff, Fernandes and Feiner \cite{Re12} proposed a subtle dynamic FOV modification strategy, by which the FOV automatically changes based on the linear and angular speed in the virtual scene. Their results suggest that it could reduce users’ VR sickness without decreasing presence in virtual navigation. Recently, Adhanom et al. \cite{Re14} upgraded the restrictor by dynamically moving its center according to the user’s eye gaze position to provide a wider visual scan area. In this research, we adopted the most widely applied implementation of dynamic FOV reduction method \cite{Re12} as one of our conditions.

The accommodation-convergence conflict occurs due to inconsistencies between VR displays and the physical reality, and this can cause VR sickness \cite{Re10,Re30}. Depth of Field (DOF) blur effects can help correct focal cues to reduce visual fatigue and improve the quality of the 3D experience \cite{Re19}. It can lessen the strain caused by the accommodation-convergence conflict. Hillaire et al. \cite{Re20,Re31} added a temporal filter to the final focus-distance computation to make the DOF blur effects and found that these degraded the participants’ performance in a first-person shooting game. Carnegie and Rhee \cite{Re18} implemented a real-time dynamic DOF technique that keeps the center of the screen in focus and reported that the usage of DOF blurring could reduce visual discomfort. However, they also concluded that their proposed blur effect was not always useful for reducing visual discomfort in HMDs. DOF blurring is another condition we used in our study.

A rest frame (RF, or frame of reference) has been proposed as a simple yet effective method for reducing VR sickness \cite{Re25}. It refers to a specific, fixed reference frame to help reduce the sensory mismatch of users \cite{Re32}. Prior studies have tried different types of RFs. For instance, Cao et al. \cite{Re33} included a black metal net in the VE. They found that the net alleviated the discomfort compared to the VE without a RF. However, some designs, such as orange spheres on the sky \cite{Re34}, a desk \cite{Re35} or cockpit and a radial together \cite{Re26}, did not assist in reducing VR sickness in specific experiments. Other studies \cite{Re17,Re21,Re22,Re23} added a virtual nose to the middle bottom part of the view to reduce VR sickness. Rather than using a virtual nose, Clark et al. \cite{Re24} added a target reticule in the center of the screen in a first-person perspective Pac-Man game. They reported that the game version with target reticule reduced users’ feelings of dizziness and nausea. A target reticule is commonly used in first-person shooting games to provide a fixed focal point. This fixed point is helpful for players to focus on while there are other rapidly moving targets, and for reducing players’ feelings of motion sickness \cite{Re36}. In this study, we investigate the effect of adding a target reticule because it does not restrict the visual space and can be applied to more scenarios compared to other designs of RF. 

The degree of control the user has in VEs also influences sickness levels because it allows one to anticipate future motions so that the conflict can be alleviated \cite{Re4,Re7,Re9,Re30}. This is similar to the common experience that passengers are more possibly to have negative symptoms than drivers who have control of the car. This finding has also been verified in desktop VEs \cite{Re37}. When the interactivity is high in VE, the users may have greater involvement and enjoyment, and instead, less sickness symptoms \cite{Re11}. However, a recent study \cite{Re38} showed opposite results in a VR driving simulation. Given the results from the limited available research, we considered the degree of control as one of the factors and investigate its effects.

In summary, previous work has presented different visual reduction techniques for eliminating or minimizing VR sickness. In this study, we aimed at comparing in a systematic manner the effects of the sickness reduction methods in a first-perspective racing game that requires rapid movements and understanding the impact of users’ control.

\section{User Study}
In this section, we describe the experiment conducted in this research. The aim of this user study was to compare the effects of VR sickness mitigation methods in terms of VR sickness, presence, workload to complete a search task, and information loss with two levels of control. 

\subsection{Participants and Apparatus}
This user study involved 32 unpaid participants (4 females, aged from 18 to 24; mean = 19.56 ± 1.41). All of them were undergraduate or graduate students from a local university and were recruited through a social media platform. Before the experiment, participants were asked to rate their familiarity with VR devices, experience with racing games, and perceived susceptibility to VR sickness. Table~\ref{tab:participants} summarizes these results.

\begin{table*}
  \caption{Summary of participants’ ratings of their familiarity with VR devices, experience with racing games, and perceived susceptibility to VR sickness on a scale of 1 (not at all) to 5 (very much so). The ratings are provided in percentage (and count).}
  \label{tab:participants}
  \begin{tabular}{cccccl}
    \toprule
    Items & 1 & 2 & 3 & 4 & 5\\
    \midrule
    Familiarity with VR devices & 6\% (2) & 19\% (6) & 44\% (14) & 19\% (6) & 13\% (4)\\
    Experience with racing games & 6\% (2) & 19\% (6) & 31\% (10) & 25\% (8) & 19\% (6)\\
    Perceived susceptibility to VR sickness & 34\% (11) & 44\% (14) & 9\% (3) & 9\% (3) & 3\% (1)\\
    \bottomrule
  \end{tabular}
\end{table*}

\begin{figure}
    \includegraphics[width=\linewidth]{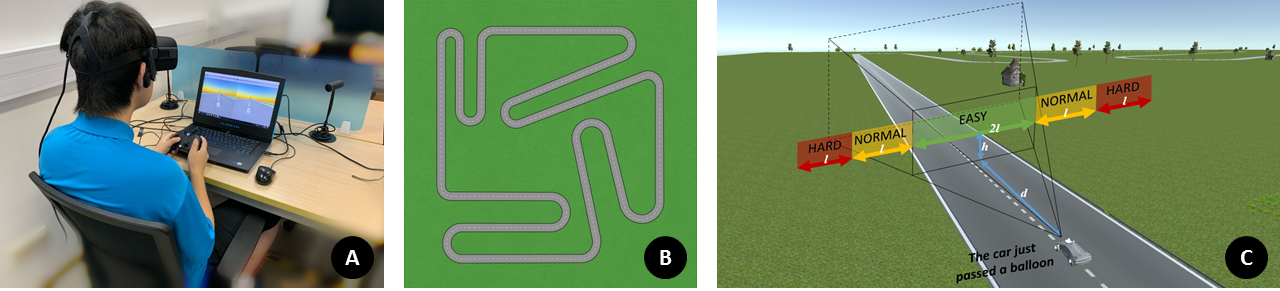}
    \caption{(A) A participant wearing an Oculus HMD and holding a gamepad controller; (B) A bird’s eye view of the racetrack; and (C) Difficulty levels of balloons based on their location in relation to the user's view.}
    \label{fig:exp}
    \Description{figure1}
\end{figure}

We used an Oculus Rift CV1 as our VR HMD. It was connected to a laptop with an Intel i9 processor and GTX 1080 GPU. An Xbox 360 controller was used to allow participants the control of the environment. Participants were requested to sit on a chair during the experiment (Figure~\ref{fig:exp}A). Sitting reduces the demand on postural control, and it is safer and more comfortable than standing \cite{Re4}. We used the Unity3D engine to develop a race game (which run at 60 fps) for this experiment, which is discussed in the following section. 

\subsection{Virtual Environment and Setting}
The Easy Roads3D asset\footnote{https://assetstore.unity.com/packages/tools/terrain/easyroads3d-pro-v3-469} from Unity Asset Store was adapted to build the experimental environment. The racetrack included 8 right turns and 4 left turns (Figure~\ref{fig:exp}B). There were 57 pairs of arrow signs on the road to show the directions to participants and 12 directional signs placed by the roadside to warn them of any upcoming turns. Guardrails were placed on both sides of the road. In order to make the virtual scene more realistic, we added other objects, including grass, trees, and wooden houses to the VE. The participants would use a joystick controller to drive a virtual car or just sit in the car without control similar to sitting in a roller-coaster (more on this later) in first-person perspective. For experimental purposes, the body of the virtual car was set to be invisible because it could be an extra RF for participants which would represent an unwanted confounding factor. 

Sixty red balloons were placed at different places floating in the air. Participants had to press the trigger on the controller to confirm they have seen the balloons. To avoid the case where multiple balloons may appear in the same view, only one balloon could be present—that is, only if participants passed the location of a balloon and it was behind their view, the next balloon would appear. The height (\textit{h}) between balloons and the ground was fixed in the range of 4 to 6 units of length and the pavement distance (\textit{d}) between two balloons was fixed to 147 units. To understand the level of information loss, we categorized balloons into three difficulty levels (20 in each level) based on their location (see Figure~\ref{fig:exp}C): (1) Easy. The balloon would appear in the range of FOV with the width of \textit{2l} that was \textit{d} apart from the current position (i.e., the car has just passed the previous balloon). (2) Normal. The balloons would appear in the area with half of the horizontal distance in Easy (\textit{l}) in both sides. (3) Hard. The balloons would appear within \textit{l} but outside the Normal level view. The participants \textit{do not need}, \textit{slightly need}, and \textit{necessarily need} to move their head to change the horizontal view to find the balloons in the three levels, respectively.

\begin{figure}
    \includegraphics[width=\linewidth]{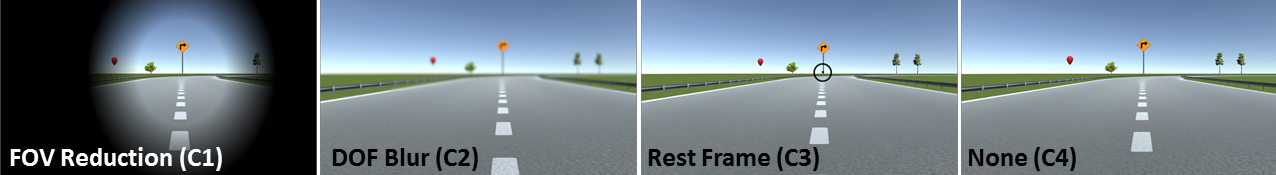}
    \caption{Participants’ view in each condition: FOV Reduction (C1), DOF Blur (C2), Rest Frame (C3), and None (C4).}
    \label{fig:conditions}
    \Description{four conditions}
\end{figure}

We generated the following four conditions within the VE to compare the effects of the previously mentioned techniques (as depicted in Figure~\ref{fig:conditions}): 
\begin{itemize}
\item {\textbf{FOV Reduction (C1)}}: In this condition, we applied the dynamic FOV reduction strategy, tunneling\footnote{https://assetstore.unity.com/packages/tools/camera/vr-tunnelling-pro-106782}, as discussed in \cite{Re12,Re13,Re14}. We used a black texture with a transparent circular cut-off. The FOV would change according to the linear and angular velocity \cite{Re12} of the virtual car, i.e., the FOV is restricted when the car speeds up or changes directions.
\item {\textbf{DOF Blur (C2)}}: We used the Post-processing package\footnote{https://github.com/Unity-Technologies/PostProcessing} to realize the DOF blur effects. It simulated the focus properties of a camera lens. Objects outside the distance to the point of focus were blurred. The blurring effects would only appear while the car was moving.
\item {\textbf{Rest Frame (C3)}}: Like \cite{Re24}, we added a target reticule for this condition. The reticule was placed in the center of the scene and followed the users’ camera. We chose the reticule as the RF because it has shown to be effective and is a minimalistic design when compared to other RFs.
\item {\textbf{None (C4)}}: C4 works as a baseline condition in which no techniques were applied.
\end{itemize}

\subsection{Design, Task and Procedure}
Our study used a 2 × 4 mixed factorial design with user’s degree of control (\textit{Control}) as a between-subjects factor and the visual methods (\textit{Technique}) as a within-subjects factor. \textit{Control} had two conditions: (1) high degree of control (HDoC), in which participants needed to drive the car during the experiment; (2) low degree of control (LDoC), in which participants did not need to drive. There were equal number of participants in the above two conditions (16 each). As stated before, \textit{Technique} had four conditions: FOV reduction (C1), DOF blur (C2), Rest frame (C3), and None (C4; the baseline condition). The order of \textit{Technique} was determined using a Latin Square approach across participants to avoid carry-over effects.

Participants were asked to press the trigger on the controller to confirm that they saw a balloon while sitting in the virtual car in LDoC. While in HDoC, in addition to finding the balloons, participants were required to use the joystick to drive the virtual car as fast as possible but also trying to avoid any collisions. 

Participants were first invited to complete a general questionnaire before the experiment. Then, we briefed them about the study, equipment, VE, and tasks. At the beginning of each trial, participants had a simple practice session to help them become familiar with the controls and each technique. The map in the training session included a short straight road, a left turn and 10 balloons (3 in Easy level, 3 in Normal level, and 4 in Hard level). Once they finished a practice session, they were required to complete the tasks while running a full lap in the racetrack. A break was given between two sessions. During the break, participants were asked to complete a few questionnaires to rate their feelings in the just-finished session. There was no restriction on the duration of the break. Participants could rest as much as they wanted, until they thought they could continue with the next session without negative feelings. When they did all four \textit{Technique} conditions, the experiment was over. They could stop at any point for any reason (e.g. if they felt extremely uncomfortable) during the experiment. In LDoC, the completion time of a lap was set to 320 seconds which was the average time the we found in a preliminary, pilot study. In HDoC, each trial took approximately 5 minutes. The whole experiment took about 30 minutes without the rest periods.

\subsection{Measurements}
Four questionnaires were given to participants. At the beginning of the experiment, a general questionnaire was used to collect participants’ demographic information and their familiarity with VR devices and racing games, and their perceived susceptibility to VR sickness. We used the Slater-Usoh-Steed Presence Questionnaire (PQ) \cite{Re39} to elicit participants’ presence and the NASA TLX Questionnaire \cite{Re40} to investigate the workload for each trial. Additionally, the Simulator Sickness Questionnaire (SSQ) \cite{Re41} was used to measure the level of sickness. According to \cite{Re41}, the results of SSQ can be grouped into four scores: Nausea (N), Oculomotor (O), Disorientation (D), and Total Severity (TS). SSQ was given to participants before the experiment and after each session \cite{Re41}. We considered the relative SSQ scores used in many existed research (e.g. \cite{Re13,Re34}), i.e., any changes in the level of sickness before and after the trial ($\Delta$SSQ), as the measurement for a certain session. The first SSQ, as a baseline of participants’ symptoms, was attached at the end of the general questionnaire before the formal experiment began. 

We collected participants’ performance by recording the number of balloons they confirmed in each difficulty level. The unconfirmed balloons were further grouped into two categories: (1) missed or (2) unseen. Missed balloons were balloons that appeared in participants’ FOV but were not confirmed by the participants. Unseen balloons were those that did not appear in participants’ FOV. Besides, we also collected the completion time and the number of collisions with guardrails in the HDoC.

\subsection{Hypotheses}
This user study is aimed at verifying the following hypotheses that are derived from the literature:
\begin{itemize}
    \item {\textbf{\textit{H1}}}: Participants will have more VR sickness in LDoC than in HDoC.
    \item {\textbf{\textit{H2}}}: The techniques applied in C1, C2, and C3 will reduce the VR sickness compared to the baseline (C4).
    \item {\textbf{\textit{H3}}}: Participants will have less sense of presence in VE in C1, C2, and C3 compared to in C4. 
    \item {\textbf{\textit{H4}}}: There will be less information loss in C4 than in the other three conditions. 
\end{itemize}

\section{Results}
We used SPSS version 26 for data analysis. Two-way mixed ANOVA tests were employed with \textit{Technique} (C1, C2, C3, C4) as the within-subjects variable and \textit{Control} (HDoC and LDoC) as the between-subjects variable. Bonferroni correction was used for pairwise comparison if a significant difference was found in the ANOVA test. Table~\ref{tab:ANOVA} shows the test results for the measures of VR Sickness, presence, workload to complete the tasks, and information loss. In addition, Table~\ref{tab:descriptive} summarizes the means and standard deviations of these measures. 

\begin{table*}
  \caption{The two-way mixed ANOVA test results for the measures of Simulator Sickness Questionnaire, SUS Presence Questionnaire, NASA TLX Questionnaire, and information loss. Significant values are in bold and marked with superscript (*, **, and *** highlights \textit{p} < .05, < .01, and < .001, respectively).}
  \label{tab:ANOVA}
  \begin{tabular}{>{\centering\arraybackslash}p{0.12\linewidth}
  >{\centering\arraybackslash}p{0.07\linewidth}
  >{\centering\arraybackslash}p{0.09\linewidth}
  >{\centering\arraybackslash}p{0.06\linewidth}
  >{\centering\arraybackslash}p{0.07\linewidth}
  >{\centering\arraybackslash}p{0.07\linewidth}
  >{\centering\arraybackslash}p{0.06\linewidth}
  >{\centering\arraybackslash}p{0.07\linewidth}
  >{\centering\arraybackslash}p{0.07\linewidth}
  >{\centering\arraybackslash}p{0.06\linewidth}l}
    \toprule
     & \multicolumn{3}{>{\centering\arraybackslash}p{0.22\linewidth}}{\textit{Technique}} & \multicolumn{3}{>{\centering\arraybackslash}p{0.2\linewidth}}{\textit{Control}} & \multicolumn{3}{>{\centering\arraybackslash}p{0.21\linewidth}}{\textit{Technique $\times$ Control}}\\
    \hline
    & \textit{F} & \textit{p} & $\eta_p^2$ & \textit{F} & \textit{p} & $\eta_p^2$ & \textit{F} & \textit{p} & $\eta_p^2$\\
    \midrule
    \multicolumn{10}{p{\dimexpr \linewidth-2\tabcolsep} }{Simulator Sickness Questionnaire (SSQ, with 4 sub-measures, N = Nausea, O = Occulomotor, D = Disorientation, TS = Total Severity)}\\
    \hline
	SSQ-N & 3.726 & \textbf{.014*} & .110 & 2.712 & .110 & .083 & 1.211 & .310 & .039\\
	SSQ-O & 2.918 & \textbf{.038*} & .089 & 1.029 & .319 & .033 & 1.849 & .144 & .058\\
	SSQ-D & 4.153 & \textbf{.008**} & .122 & 1.420 & .243 & .045 & 1.059 & .370 & .034\\
	SSQ-TS & 4.056 & \textbf{.009**} & .119 & 1.935 & .174 & .061 & 1.557 & .205 & .049\\
	\hline
	\multicolumn{10}{p{\dimexpr \linewidth-2\tabcolsep}}{Slater-Usoh-Steed Presence Questionnaire (PQ)}\\
    \hline
    PQ & 1.296 & .281 & .041 & .044 & .836 & .001 & .356 & .785 & .012\\
    \hline
    \multicolumn{10}{p{\dimexpr \linewidth-2\tabcolsep}}{NASA TLX Questionnaire (NASA TLX)}\\
    \hline
    NASA TLX & 2.135 & .101 & .066 & .303 & .586 & .010 & 1.032 & .382 & .033\\
    \hline
    \multicolumn{10}{p{\dimexpr \linewidth-2\tabcolsep}}{Information loss (number of balloons that participants confirmed (C), missed (M), or unseen (U) in Easy, Normal, Hard level or considering all levels together (Total))}\\
    \hline
    Easy-C & 5.177 & \textbf{.002**} & .147 & .475 & .496 & .016 & .440 & .725 & .014\\
    Easy-M & 5.130 & \textbf{.003**} & .146 & 1.102 & .302 & .035 & 1.369 & .257 & .044\\
    Easy-U & 35.520 & \textbf{<.001***} & .542 & .110 & .742 & .004 & .495 & .687 & .016\\
    Normal-C & 3.540 & \textbf{.018*} & .106 & .800 & .378 & .026 & .255 & .857 & .008\\
    Normal-M & 21.173 & \textbf{<.001***} & .414 & .055 & .817 & .002 & 3.393 & \textbf{.021*} & .102\\
    Normal-U & 16.909 & \textbf{<.001***} & .360 & .851 & .364 & .028 & 1.457 & .232 & .046\\
    Hard-C & 10.241 & \textbf{<.001***} & .254 & .365 & .550 & .012 & .388 & .762 & .013\\
    Hard-M & 35.670 & \textbf{<.001***} & .543 & .416 & .524 & .014 & 1.289 & .283 & .041\\
    Hard-U & 13.866 & \textbf{<.001***} & .316 & .522 & .475 & .017 & 1.362 & .260 & .043\\
    Total-C & 6.837 & \textbf{<.001***} & .186 & .595 & .446 & .019 & .116 & .951 & .004\\
    Total-M & 43.787 & \textbf{<.001***} & .593 & .241 & .627 & .008 & .988 & .402 & .032\\
    Total-U & 41.361 & \textbf{<.001***} & .580 & .523 & .475 & .017 & 1.216 & .308 & .039\\
    \bottomrule
    
  \end{tabular}
\end{table*}


\begin{table*}
  \caption{Quantitative measures of Simulator Sickness Questionnaire (SSQ, with 4 sub-measures, N = Nausea, O = Oculomotor, D = Disorientation, TS = Total Severity), SUS Presence Questionnaire, NASA TLX Questionnaire in terms of mean (standard deviation).}
  \label{tab:descriptive}
  \begin{tabular}{
  >{\centering\arraybackslash}p{0.06\linewidth}
  >{\centering\arraybackslash}p{0.03\linewidth}
  >{\centering\arraybackslash}p{0.1\linewidth}
  >{\centering\arraybackslash}p{0.1\linewidth}
  >{\centering\arraybackslash}p{0.1\linewidth}
  >{\centering\arraybackslash}p{0.1\linewidth}
  >{\centering\arraybackslash}p{0.1\linewidth}
  >{\centering\arraybackslash}p{0.1\linewidth}}
    \toprule
    && SSQ-N & SSQ-O & SSQ-D & SSQ-TS & PQ & NASA TLX\\
    \midrule
    HDoC&C1&4.17 (21.75)&10.42 (17.91)&5.22 (36.25)&8.18 (25.19)&23.69 (3.80)&41.29 (18.34)\\
	&C2&11.33 (17.84)&9.00 (18.82)&26.97 (30.27)&16.13 (20.01)&22.94 (5.26)&45.29 (19.36)\\
	&C3&0.00 (13.93)&-0.47 (11.23)&2.61 (25.01)&0.47 (16.15)&24.00 (3.66)&38.08 (15.39)\\
	&C4&2.39 (14.57)&1.90 (11.24)&4.35 (28.63)&3.04 (14.14)&23.88 (5.59)&40.94 (15.8)\\
	\hline
    LDoC&C1&-3.58 (30.74)&-2.37 (28.05)&-14.79 (65.38)&-6.55 (40.01)&24.59 (7.81)&39.44 (21.56)\\
	&C2&28.02 (41.73)&27.00 (43.94)&53.07 (87.92)&38.57 (60.18)&22.19 (7.43)&38.94 (24.31)\\
	&C3&5.37 (15.95)&0.00 (22.66)&5.22 (37.99)&3.51 (24.41)&23.20 (6.99)&37.56 (15.43)\\
	&C4&6.56 (18.01)&6.16 (19.81)&18.27 (41.83)&10.52 (24.32)&22.69 (8.14)&35.85 (19.29)\\
    \bottomrule
  \end{tabular}
\end{table*}

As mentioned in Section 3.4, we calculated relative SSQ scores with four sub-measures: SSQ-N, SSQ-O, SSQ-D, and SSQ-TS. The results of ANOVA tests showed that there was no significant interaction effect between \textit{Technique} and \textit{Control} for any of the SSQ measures (\textit{p} > .05, see Table~\ref{tab:ANOVA}). Likewise, no significant difference between \textit{Control} was found with respect to all SSQ scores (\textit{p} > .05, see Table~\ref{tab:ANOVA}). However, significant main effects among \textit{Technique} conditions were found with respect to SSQ-N (\textit{F}(3, 90) = 3.726, \textit{p} = .014, $\eta_p^2$ = .110), SSQ-O (\textit{F}(3, 90)= 2.918, \textit{p} = .038, $\eta_p^2$ = .089), SSQ-D (\textit{F}(3, 90)= 4.153, \textit{p} = .008, $\eta_p^2$ = .122), and SSQ-TS (\textit{F}(3, 90)= 4.056, \textit{p} = .009, $\eta_p^2$ = .119). The post-hoc analysis using Bonferroni-adjusted paired t-tests only showed a significant difference between C2 and C3 in SSQ-D (\textit{p} = .037), while the differences in the remaining pairwise comparisons were not significant (\textit{p} > .05). Results of Spearman's rank-order test showed that there were significant positive correlations between perceived susceptibility to VR sickness and either average SSQ-N score (\textit{$r_s$} = .526, \textit{N} = 16, \textit{p} = .036) or average SSQ-O score (\textit{$r_s$} = .532, \textit{N} = 16, \textit{p} = .034) of four \textit{Technique} conditions in HDoC. However, no significant associations were found between the remaining combinations in HDoC or in LDoC. 

The presence of being in the VE was measured by PQ scores. ANOVA test did not yield a significant interaction effect between \textit{Technique} and \textit{Control} (\textit{F}(3, 90)= .356, \textit{p} = .785, $\eta_p^2$ = .012). In addition, there was no significant main effect of \textit{Technique} (\textit{F}(3, 90)= 1.296, \textit{p} = .281, $\eta_p^2$ = .041) and of \textit{Control} (\textit{F}(1, 30)= .044, \textit{p} = .836, $\eta_p^2$ = .001) on PQ scores. We performed a Spearman's rank-order correlation test and only found a positive correlation between participants' ratings in familiarity to racing game and their ratings in PQ scores in LDoC condition that was significant (\textit{$r_s$} = .642, \textit{N} = 16, \textit{p} = .007).

We computed the overall workload based on the ratings and weights given by participants for the NASA TLX scores. The results of the ANOVA test showed that there was also no significant interaction effect between \textit{Technique} and \textit{Control} (\textit{F}(3, 90)= 1.032, \textit{p} = .382, $\eta_p^2$ = .033). In addition, there was no significant main effect of \textit{Technique} (\textit{F}(3, 90)= 2.135, \textit{p} = .101, $\eta_p^2$ = .066) and \textit{Control} (\textit{F}(1, 30)= .303, \textit{p} = .586, $\eta_p^2$ = .010) on workload scores. Results from Spearman's rank-order correlation test showed no significant correlations between participants' self-evaluation ratings and their reported workloads in both levels of \textit{Control} (\textit{p} > .05).

\begin{figure}
  \includegraphics[width=\textwidth]{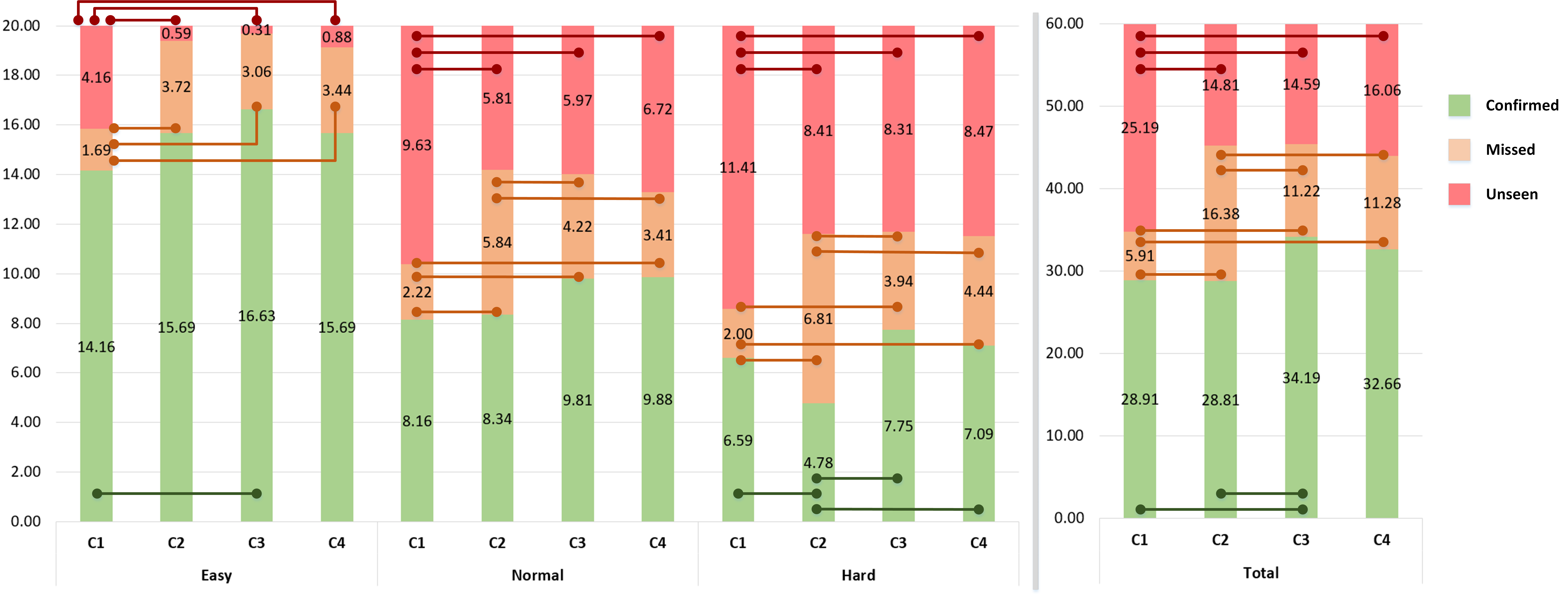}
  \caption{Average number of balloons that participants confirmed, missed or unseen under the four conditions. Horizontal bars represent that there is a significant difference between two conditions according to Bonferroni adjusted pairwise comparisons at \textit{p} < .05 level.}
  \label{fig:balloonResults}
  \Description{Average number of balloons}
\end{figure}

The two-way mixed ANOVA test revealed that the interaction effect between \textit{Technique} and \textit{Control} was only found in the number of missed balloons in Normal level (\textit{F}(3, 90)= 3.393, \textit{p} = .021, $\eta_p^2$ = .102; \textit{p} > .05 for the remaining, see Table~\ref{tab:ANOVA}). We found significant effects of \textit{Technique} condition on the number of balloons regardless of their difficulty levels (\textit{p} < .05, see Table~\ref{tab:ANOVA} for details). Figure~\ref{fig:balloonResults} depicts the average number of balloons across all four \textit{Technique} conditions. The significant differences found in Bonferroni pairwise comparisons are also highlighted in Figure~\ref{fig:balloonResults}. On the contrary, we did not find significant difference between \textit{Control} with respect to the number of confirmed, missed, or unseen balloons under Easy, Normal, Hard levels, or considering all levels together (\textit{p} > .05, see Table~\ref{tab:ANOVA}). We also conducted a Spearman's rank-order test to examine the correlation between participants' self-evaluation ratings before the experiment and their overall information loss. In HDoC, there was a significant negative correlation between the number of confirmed balloons and participants' ratings in their perceived susceptibility to VR sickness in HDoC (\textit{$r_s$} = -.610, \textit{N} = 16, \textit{p} = .012) but no significant relationships between participants' overall information loss and neither their ratings in familiarity to VR devices nor their familiarity to racing games was found. In LDoC, no significant correlations were found (\textit{p} > .05).

A Pearson correlation test was conducted to investigate the relationships between SSQ-TS scores and total number of balloons (i.e., sum all three levels together) in each \textit{Technique} and \textit{Control} condition. Table~\ref{tab:corr} shows the correlation matrix. The significant correlations were all found in C2. In HDoC, there was a positive relationship between SSQ-TS scores and the number of unseen balloons (\textit{r} = .628, \textit{N} = 16, \textit{p} = .009). In LDoC, there was a negative association between SSQ-TS scores and the number of confirmed balloons (\textit{r} = -.601, \textit{N} = 16, \textit{p} = .014) and a positive association between SSQ-TS scores and the number of unseen balloons (\textit{r} = .714, \textit{N} = 16, \textit{p} = .002).

\begin{table}
  \caption{Correlation matrix among SSQ-TS scores and total number of balloons for each condition (\textit{N} = 16 for each entry). C = Confirmed, M = Missed, U = Unseen. Significant values are in bold. Correlation is significant at the .05 level (*) or at the .01 level (**).}
  \label{tab:corr}
  \begin{tabular}{ccccccccccl}
    \toprule
    &&\multicolumn{4}{c}{HDoC}&\multicolumn{4}{c}{LDoC}\\
    \hline
    &&C1&C2&C3&C4&C1&C2&C3&C4\\
    \midrule
    C1&C&.051&&&&.461&&&\\
    &M&.408&&&&.171&&&\\
    &U&-.172&&&&-.492&&&\\
    \hline
    C2&C&&-.368&&&&\textbf{-.601*}&&\\
    &M&&-.248&&&&.023&&\\
    &U&&\textbf{.628**}&&&&\textbf{.714**}&&\\
    \hline
    C3&C&&&-.064&&&&.017&\\
    &M&&&-.187&&&&.074&\\
    &U&&&.183&&&&-.057&\\
    \hline
    C4&C&&&&-.079&&&&.066\\
    &M&&&&.126&&&&-.141\\
    &U&&&&.040&&&&-.033\\

    \bottomrule
  \end{tabular}
\end{table}

A one-way ANOVA test found that there were no significant difference in completion time (\textit{F}(3,60) = .563, \textit{p} = .642) and number of collisions (\textit{F}(3,60) = .193, \textit{p} = .901) among the four \textit{Technique} conditions in HDoC. The overall average completion time in HDoC was very close to our setting in LDoC (320s). Results from the Pearson correlation test found no significant association between completion time and number of collisions (\textit{r} = -.143, \textit{N} = 64, \textit{p} = .260).

\section{Discussion}
Contrary to our expectation, there were no significant differences on VR sickness between HDoC and LDoC (\textbf{\textit{H1}} could not be confirmed). Overall, the results did not confirm our hypothesis \textbf{\textit{H2}}. Results showed that applying visual reduction techniques, whether it is reducing FOV, applying DOF blur effects or having a target reticule as a rest frame did not significantly reduce the sickness level in relation to the baseline. These findings are different from previous work \cite{Re12,Re13,Re21,Re22,Re24,Re33}. One main reason might be due to the difference of the virtual scenarios and the given tasks. These prior studies focused on virtual navigation in which their participants were required to control the motion to walk around in the VEs. Compared to such navigating process, our racing game involved higher self-motion speed and higher rate of linear or rotational acceleration, which would have likely led to greater level of VR sickness \cite{Re4,Re8,Re30}. As such, we can provisionally conclude that FOV reduction, DOF blurring, and target reticule as a rest frame cannot efficiently help reduce the sickness in VR scenarios involving high speed movements, such as a racing game.

After experiencing the FOV reduction method (C1) in HDoC, some participants reported that they felt discomfort and disoriented because of the frequent changes in the size of their visual view, especially when a collision occurred. While we did not receive related feedback from the participants in LDoC but compared to the self-controlled driving speed in HDoC, the speed fluctuated less throughout in LDoC, thus causing fewer modifications in FOV in LDoC. It is possible that the frequent changes in FOV caused by the rapid changes in self-motion speed could lead to negative symptoms. Further investigations with user studies using various scenarios are required to explore this possibility. 

As shown in Table~\ref{tab:descriptive}, participants reported strong negative feelings in C2. Meanwhile, there were significantly more missed balloons in C2 compared to other \textit{Technique} conditions (Figure~\ref{fig:balloonResults}, in terms of total number of balloons). However, there was no significant correlation between SSQ-TS scores and the total number of missed balloons in this condition. Instead, the significant relationship was found in SSQ-TS scores and the total number of unseen balloons (see Table~\ref{tab:corr}). Possibly, the participants have felt discomfort due to the blur effect, and tried to reduce their head movements to avoid increasing the sickness level. One participant in LDoC [\textit{P14}] commented that "\textit{The blur effect made me uncomfortable and it was hard for me to find the balloons}". According to \cite{Re30}, the blur level in the scene needs to be subtle enough to match the depth effects as they occur in real life. We suggest a careful consideration may be needed before adding DOF blurring into a VE; if the blur levels used are not appropriate, it could possibly increase VR sickness instead of the other way around. 

Relative SSQ scores in C3 were lower than in C4 in both \textit{Control} conditions, but the differences were not noticeable. A target reticule may have been too small to achieve the significant reduction effect in VR sickness in a fast pace environment like the racing game used in this study. The size of the RF represents a tradeoff between a clearer, unobstructed view and VR sickness. Designers can consider using a rest frame with a suitable size and have a proper reason for its appearance.

The results did not support \textbf{\textit{H3}}. The presence questionnaire scores were close in the two \textit{Control} conditions. Participants did not have a weaker sense of presence when using these techniques. This could be caused by the scenario we used, in which they were required to find targets in a fast pace environment in a short time. 

Based on the results of NASA TLX, the overall average workload in HDoC was higher than in LDoC under all four \textit{Technique} conditions though the differences were not significant. One possible reason for the lack of significance may be because the primary task for participants is the same in both \textit{Control} conditions, which is to find and confirm seen balloons. Besides, it is possible that the design of the racetrack was somewhat straightforward as such participants might have been able to anticipate to a certain degree the future motion and change of view, even though they were not controlling the car movements. A more complex driving scenario, possibly having intersections in the game, can be used for further investigation.

There was no significant difference in information loss between HDoC and LDoC. Participants were able to search for the objects around the environment and steer the car simultaneously. The results from statistical tests partially supported \textbf{\textit{H4}}. First, applying the technique in C1, a FOV restrictor, would result in significantly more unseen balloons in all three levels of difficulty. During the experiments, we observed that participants moved their head more frequently in C1 compared to other conditions in order to search for the balloons. A FOV restrictor decreases users’ visible range, producing a somewhat severe information loss. Unlike participants who did not see the balloons in C1, participants in C2 missed the balloons in Normal and Hard levels. The missed balloons can represent to a certain extent that the participants have seen the balloons but cannot clearly identify them. Our results indicate that DOF blur effects would lead to information loss because this technique induces a higher difficulty for users to recognize the objects. The number of missed balloons under Easy level was comparable between C2 and either C3 or C4. One possible explanation for this is that such objects would be able to enter the sharp area and can be easily identified by the participants. Third, there was no significant difference in information loss between C3 and C4 (baseline). As such, it would seem that using a target reticule would not result in information loss.

Based on the results of the experiment with the racing game, we can extrapolate the following suggestions:
\begin{itemize}
    \item Dynamic FOV reduction does not seem to work well in scenarios which involve frequent changes in linear or angular acceleration because in such scenarios, as observed from the results in this study, FOV would change often and can generate additional negative feelings for players. In addition, reducing FOV would lead to a certain degree of information loss in VEs. 
    \item When using DOF blur effects, it may be useful to adjust them to match the visual experience similar to real life. Otherwise, an excessive blur level would not only hinder participants from collecting information from VEs, but could also increase VR sickness.  
    \item A small size of rest frame (target reticule) may not be able to help reduce VR sickness significantly. However, RFs with large size may restrict the design space and reduce the level of immersion in VEs. A suitable RF should have a reasonable size and appearance to keep the balance between reducing VR sickness while maintaining (or even improving) the immersive gameplay experience. 
\end{itemize}

The main limitation in our study is that the experiment only involved a single population (though it still represents one of the most common users of VR and racing games). Individual factors, including age, gender, and illness, can have an impact on the level of VR sickness \cite{Re3,Re7,Re9}. Individual difference is also a cause of the large variances that occurred in our results. While it has been shown that people become less susceptible to VR sickness with age \cite{Re4}, it is necessary to do further explorations on the effects of these techniques in different age groups. In addition, although gender difference is not the focus in our study, it would be useful if future experiments have a balance between female and male participants. In this study, we did not see salient patterns of the relationships between participants' self-evaluation ratings before the experiments and post-experimental measurements. Further research with larger and more diverse populations may help to improve our understanding of any relationships. 

In this work, we focused on the subjective measurements of VR sickness. Future work can also adopt other monitoring methods to include physiological data like electroencephalography (EEG) to analyze visual and virtual discomfort \cite{ReJL}. However, wearing a monitor device together with HMD simultaneously may affect the accuracy of data. Other studies have also reported issues with collecting and using physiological data (e.g., \cite{Re9,Re22}). If these issues can be overcome, it will be useful to include quantitative data in experiments. As mentioned before, we can conduct the tests in different tasks and VEs. As different VR content and scenarios can have a direct influence on discomfort \cite{Re10}, it is valuable to examine the effects of these techniques in various scenarios. 

\section{Conclusion}
In this paper, we conducted a mixed factorial user study to compare the effects of different virtual reality (VR) sickness mitigation techniques, including Field of View (FOV) reduction, Depth of Field (DOF) blurring, and adding a target reticule as a rest frame, on VR sickness, presence, information loss and workload under two different levels of control in a VR racing game. Our results show no significant differences in VR sickness, presence, and workload among these techniques and under the two control levels in the VR racing game. We found that both FOV reduction and DOF blurring techniques resulted in information loss to some extent. We found that some participants felt discomfort when FOV changed frequently due to changes in their steering speed. Further research can be conducted to explore the impact of the frequency of FOV modifications with various VEs, scenarios, and tasks to help develop prescriptive guidelines for their provision. Similar, a proper blur level needs to be determined before applying DOF blur effects to a VE. While this research represents a first attempt to examine systematically the relative effects of these visual mitigation techniques, future work using different VEs, scenarios, and tasks is still needed and useful to compare their effects and to further understand how we can best leverage them to help minimize VR sickness while minimizing information loss and maximizing gameplay experience.

\begin{acks}
The authors would like to thank the participants for their time and the reviewers for their feedback. The work is supported in part by Xi’an Jiaotong-Liverpool University (XJTLU) Key Special Fund (KSF-A-03) and XJTLU Research Development Fund.
\end{acks}

\bibliographystyle{ACM-Reference-Format}
\bibliography{ref}


\end{document}